\documentclass[preprint,trackchanges]{aastex631}
\hypersetup{linkcolor=blue,citecolor=blue,filecolor=cyan,urlcolor=blue}
\shorttitle{Filament Formation}
\shortauthors{Yang et al.}

\begin{document}

\title{Formation of a Solar Filament by Magnetic Reconnection and Associated Chromospheric Evaporation and Subsequent Coronal Condensation}

\correspondingauthor{Bo Yang}
\email{boyang@ynao.ac.cn}

\author{Bo Yang}
\affil{ Yunnan Observatories, Chinese Academy of Sciences, 396 Yangfangwang, Guandu District, Kunming, 650216, People's Republic of China}
\affiliation{Center for Astronomical Mega-Science, Chinese Academy of Sciences, 20A Datun Road, Chaoyang District, Beijing, 100012, People's Republic of China}
\affiliation {Key Laboratory of Solar Activity, National Astronomical Observatories of Chinese Academy of Science, Beijing, 100012, People's Republic of China }

\author{Jiayan Yang}
\affil{ Yunnan Observatories, Chinese Academy of Sciences, 396 Yangfangwang, Guandu District, Kunming, 650216, People's Republic of China}
\affiliation{Center for Astronomical Mega-Science, Chinese Academy of Sciences, 20A Datun Road, Chaoyang District, Beijing, 100012, People's Republic of China}

\author{Yi Bi}
\affil{ Yunnan Observatories, Chinese Academy of Sciences, 396 Yangfangwang, Guandu District, Kunming, 650216, People's Republic of China}
\affiliation{Center for Astronomical Mega-Science, Chinese Academy of Sciences, 20A Datun Road, Chaoyang District, Beijing, 100012, People's Republic of China}

\author{Junchao Hong}
\affil{ Yunnan Observatories, Chinese Academy of Sciences, 396 Yangfangwang, Guandu District, Kunming, 650216, People's Republic of China}
\affiliation{Center for Astronomical Mega-Science, Chinese Academy of Sciences, 20A Datun Road, Chaoyang District, Beijing, 100012, People's Republic of China}

\author{Zhe Xu}
\affil{Purple Mountain Observatory, Chinese Academy of Sciences, No.8 Yuanhua Road, Qixia District, Nanjing 210034, People’s Republic of China }



\begin{abstract}
We present the first observation of a solar filament formed by magnetic reconnection and associated chromospheric evaporation and subsequent coronal condensation.
Driven by shearing motion during flux emergence, a sequential tether-cutting reconnection processes
occurred and resulted in an M1.3 confined flare accompany with the formation of a sigmoid structure.
It is found that the flare had a conjugate compact footpoint brightenings, which correspond to the footpoints of the sigmoid.
Furthermore, observational evidence of explosive evaporation is well diagnosed at the conjugate footpoint brightenings in the impulsive phase of the flare.
After the flare, continuous cool condensations formed at about the middle section of the sigmoid
and then moved in opposite directions along the sigmoid, eventually leading to the formation of the filament.
These observations suggest that magnetic reconnection not only can form the magnetic field structure of the filament,
but also heat their chromospheric footpoints during their formation and drive chromospheric evaporation.
As a result, the heated chromospheric plasma may be evaporated into the magnetic field structure of the filament,
where the accumulated hot plasma might suffer from thermal instability or thermal non-equilibrium, causing catastrophic cooling
and coronal condensation to form the cool dense material of the filament.
This observation lends strong support to the evaporative-condensation model and highlights the crucial role of magnetic reconnection
in forming both the magnetic field structure and the cool dense material of filaments.
\end{abstract}


\section{Introduction} \label{sec:intro}
Solar filaments are filled with cool dense plasma, manifesting as cool clouds suspended in the surrounding hot tenuous corona. They appear as elongated absorption features
on the solar disk in $H_{\alpha}$ and some extreme ultraviolet (EUV) lines \citep{anz05}. Outside the solar disk, they appear as emitting plasma structures
called ``prominences". Filaments are usually formed above the polarity inversion lines (PILs),
which separate positive and negative magnetic flux regions in the photosphere \citep{mar98}. It has become a consensus that magnetic field supports
the cold and dense plasma of filaments against gravity in the corona. Despite over a century of increasingly detailed observations and studies,
the mechanism accounting for the filament formation is still controversial. Unravelling the mechanism of filament formation
requires a thorough understanding of the formation of the magnetic structure and the cool dense plasma of filaments.

As stressed by \citet{chen20a} that a filament should be a coronal structure. Many direct measurements of prominence magnetic fields
have shed some light on the magnetic structure of prominences \citep{ler89,oro14,schm14}.
However, it is difficult to determine prominence magnetic fields due to the ambiguity of Stokes inversion or the coronal magnetic fields surrounding the prominences.
The detailed magnetic structure of filaments is still far from being fully understood. Nevertheless,
previous theoretical and observational studies on filaments have deepened our understanding of the magnetic structure of filaments.
Sheared arcades \citep{ks57,dev00} and flux ropes \citep{kr74,van89} are believed to be suitable for supporting filaments.
A flux rope naturally contains magnetic dips above the PILs, and the cool dense plasma of the filaments is preferentially supported in the dips.
A sheared arcade can contain or not contain magnetic dips. Considering a filament is a dynamic entity, \citet{kar01}
suggested that filaments can even be supported by sheared arcades without magnetic dips. Both sheared arcades and flux ropes
can be formed by the surface mechanisms or the subsurface mechanisms \citep[see the review by][]{mac10}.
In the surface mechanisms \citep{van89,mar01}, magnetic reconnection occurs between a series of magnetic arcades
straddling the PILs to form the sheared arcades or flux ropes, under the joint of photospheric shear flows parallel to the PILs and converging flows perpendicular to the PILs.
In the subsurface mechanisms, a preexisting flux rope in the convection zone partly emerges through the solar surface into the corona by buoyancy \citep{rus94,fan01}.
Hitherto, a great deal of observations have tended to support the surface mechanisms \citep{chae01,yan16,yang15,yang16,yang19a,chen20b},
and only a handful of observations have been in favor of the subsurface mechanisms \citep{oka08,lit10,xu12,yan17}.

It is generally accepted that the filament plasma originates from the solar chromosphere \citep{song17,chen20a},
but the mechanism by which the chromospheric plasma is transported into the corona to form filaments is still under debate.
Historically, there are three promising models for chromospheric plasma being transported into the corona to form filaments \citep{mac10,chen20a}.
The first is the injection model, which demonstrated that chromospheric plasma can be injected into a filament channel through magnetic reconnection \citep{wang99,liu05,wang18,shen19,yang19b,wei20}.
The second is the levitation model, which suggested that chromospheric plasma can be directly lifted to the corona by emerging flux or magnetic flux cancellation \citep{rus94,zhao17}.
The third is the evaporation-condensation model, which proposed that chromospheric plasma can be heated to several million kelvin and evaporated into the corona,
and then thermal instability \citep{par53,fie65} or thermal non-equilibrium \citep{ant91,ant00} causes catastrophic cooling and coronal condensation.
The essence of this model is that artificial localized heating concentrated exclusively at the chromospheric footpoints of magnetic loops
are needed to heat the chromosphere and then drive chromospheric evaporation. More recently,
\citet{huang21} tried to unify the injection model and the evaporation-condensation model in a single framework.
They found that when the localized heating is situated in the upper chromosphere,
the local plasma is heated to evaporate into the corona; when the localized heating is situated in the lower chromosphere,
the enhanced gas pressure pushes the cold upper chromospheric material to be injected into the corona.
On the basis of the evaporation-condensation model, an increasing number of numerical simulations have successfully simulated the formation of filaments with steady or
non-steady localized heating symmetrically or asymmetrically distributed at the two chromospheric ends of filaments \citep{kar08,xia11,luna12,kep14,xia16,zhou20}.
Moreover, many dynamic phenomena of filaments, such as filament oscillations \citep[see the review by][]{chen20a}, the moving blobs \citep{luna12,xia16},
the plasma circulation of long-lived filaments \citep{xia16}, the cold $H_{\alpha}$ counterstreaming flows \citep{xia11,zhou20},
and the hot EUV counterstreaming flows between filament threads \citep{zhou20}, have also been simulated.

The evaporative-condensation model has been extensively and thoroughly analyzed in numerical simulations,
but complete and definitive observations that support and validate this model have not yet been reported.
So far, only two undisputed observations of coronal condensation forming prominences have been reported. \citet{liu12} presented a clear observation that coronal condensation occurs
at the magnetic dips of a transequatorial loop system and results in the formation of a cloud prominence after a confined eruption. Soon after, \citet{ber12} observed
the dynamic formation of a quiescent polar crown prominence in a coronal cavity and implied that the prominence is formed via in situ condensation of hot plasma from the coronal cavity.
In both cases, it is not clear whether the hot plasmas originate from chromospheric evaporation. The direct injection \citep{wangj07} and even magneto-thermal convection
involving emerging magnetic bubbles and plumes \citep{ber11} are all potential sources of the hot plasmas. In addition, recent numerical simulation \citep{kan17}
and observations \citep{li18,li21} also proposed that apart from artificial localized heating in the chromosphere,
magnetic reconnection in the corona can also trigger thermal instability and result in coronal condensation.
In this paper, we report the first definitive observation that magnetic reconnection and associated chromospheric evaporation followed by coronal condensation lead to the formation of a filament.
This observation may enhance our understanding on the filament formation.

\section{Observations}
The event was well-observed by the Atmospheric Imaging Assembly \citep[AIA;][]{lem12} and the Helioseismic and Magnetic Imager \citep[HMI;][]{sch12}
on board the \emph{Solar Dynamics Observatory} \citep[\emph{SDO};][]{pes12}. AIA takes the full-disk images of the Sun in seven EUV
and two ultraviolet (UV) wavelengths with a pixel size of 0$\arcsec$.6 at a high cadence of up to 12s.
Here, the Level 1.5 images centered at 304 \AA \ (\ion{He}{2}, 0.05 MK), 171 \AA\ (Fe {\sc ix}, 0.6 MK ), 193 \AA\ (Fe {\sc xii}, 1.3 MK and Fe {\sc xxiv}, 20 MK),
211 \AA \ (\ion{Fe}{14}, 2 MK), 335 \AA \ (\ion{Fe}{16}, 2.5 MK), 94 \AA\ (\ion{Fe}{18}, 7 MK), 131 \AA \ (Fe {\sc viii}, 0.6 MK and Fe {\sc xxi}, 10 MK),
and 1600 \AA \ (\ion{C}{4} + cont., 0.01 MK) were adopted to study the event.
HMI provides full-disk continuum intensity images and the vector magnetic field data with a pixel size of 0.$\arcsec$5.
The time cadence of the HMI are 45 s (for the continuum intensity images) and 720 s (for vector magnetic field data).
In addition, we used $H_{\alpha}$ images from the Global Oscillation Network Group (GONG) and the New Vacuum Solar Telescope \citep[NVST;][]{liu14},
soft X-ray (SXR) flux of flare at 1-8 \AA\ from the \emph {Geostationary Operational Environmental Satellites (GOES)},
and SXR image from the X-ray Telescope \citep[XRT;][]{gol07} on board the \emph{Hinode}  \citep{kos07} satellite. Images taken from the AIA, HMI, and GONG
were aligned by differentially rotating to the reference time of 22:30 UT on 2014 February 2.

To detect the chromospheric evaporation, the spectroscopic data from the EUV Imaging Spectrometer \citep[EIS;][]{cul07} onboard \emph{Hinode} satellite was also analyzed.
We mainly used the EIS data during the period of 21:31:12 UT and 21:39:58 UT that corresponds to the impulsive phase of the flare.
EIS observed this flare with 15 spectral windows, using a 2$\arcsec$ wide slit and 3$\arcsec$ step size with an exposure time of 5 s.
The spectra were taken at 80 positions, and it took an averaged duration of 8 minutes to scan an area of 240$\arcsec$$\times$304$\arcsec$.
Here, we concentrated on three strong emission lines, \ion{He}{2} 256.32 \AA, \ion{Fe}{15} 284.16 \AA, and \ion{Fe}{16}  262.98 \AA,
which provide temperature coverage from the chromosphere (log T$\approx$4.7) to the corona (log T$\approx$6.4).
The EIS level-0 data was performed with the standard Solar Software(SSW) eis\_prep.pro routine to correct the dark current, detector bias, hot pixels,
cosmic rays, and radiometric calibration. Meanwhile, the orbital variation and wavelength calibration were also corrected by using house keeping data \citep{kam10}.
Then, we used the standard SSW eis\_auto\_fit.pro routine with a single Gaussian model to derive spectral intensities, line widths and Doppler velocities.
To derive the reference wavelengths of the three strong emission lines,
the average line centers of \ion{He}{2}, \ion{Fe}{15}, and \ion{Fe}{16} lines for the quieter region of the EIS FOV were measured, respectively.
The Doppler velocities estimated by Doppler shift in these emission lines have a few km s$^{-1}$ uncertainty.

\section{Results}
\subsection{Overview of AR 11967 and Photospheric Magnetic Field Evolution }
On 2014 February 2, NOAA AR 11967 was located at about S$13\degr$E$04\degr$.
Figures 1({\it a})-({\it b}) illustrate the general appearance of the AR. On this day, four M-class flares occurred in this AR,
three of which had unique X-shape ribbons and occurred in a facular region of the AR (for a detailed analysis of these flares, see \citet{liu16}),
one had a typical two ribbons and took place in the central and northwestern section of the AR (enclosed by the blue box in Figure 1({\it a})).
Here, we focus on the M1.3 two ribbon flare, which started at about 21:24 UT, peaked at 22:04 UT, and ended at 22:14 UT in \emph{GOES} 1-8 \AA\ flux.
Apart from its two ribbons, the most striking feature of this flare is that the flare had a conjugate compact footpoint brightenings (Figure 1({\it c})).
The conjugate compact footpoint brightenings were more pronounced in the $H_{\alpha}$ observations (see the inserted images in Figure 2).
They were distributed on the sunspots with opposite polarities and corresponded to the footpoints of the subsequent formed filament (Figures 1({\it c}) and ({\it d})).

The central and northwestern section of the AR, where the flare of interest occurred, developed rapidly and formed a contiguous penumbral structure (Figure 1({\it a}))
due to magnetic flux emergence. The emerging magnetic flux was characterized by one negative sunspot, labeled as ``N",
and two major positive sunspots, labeled as ``P" and ``P1" (Figure 1({\it a})). There was constant flux emergence between P and N.
A significant flux of positive polarity emerged toward and merged with P, a significant flux of negative polarity emerged to the east of P
and then migrated eastward into N, forming an elongated channel of negative magnetic fluxes (Figure 1({\it b})).
For the sake of description, we call the elongated channel as the negative magnetic channel.
In this region, the photospheric magnetic field was highly sheared with respect to the PILs (Figure 1({\it b})).
It is notable from the NVST high-resolution $H_{\alpha}$ image (Figure 1({\it d})) that the emerged sunspots with opposite polarities were connected by highly sheared arch filament systems (AFS).
Figures 1({\it e})-({\it g}) show the evolution of the photospheric magnetic field before, during, and after the flare, covering part of the episodes of the flux emergence.
It is clear that there was significant flux cancellation ongoing along the main PILs of this region.
Unsigned negative flux integrated in a red rectangle (Figure 1({\it g})) is shown in Figure 1({\it h}).
This region was selected to avoid the negative magnetic flux that continuously migrated along the negative magnetic channel into N.
One can see that the unsigned negative flux persistently decreased from the beginning to the end of the observation.
Applying the inductive DAVE4VM \citep{sch08} method to the 12-minute cadence HMI vector magnetic field data, we calculated the photospheric velocity field,
which integrated over two hours between 20:00 UT and 22:00 UT and superimposed on a HMI vertical image (Figure 1({\it f})).
Remarkably, persistent northwestward photospheric flows characterize the positive sunspots P and P1,
while southeastward photospheric flows dominate the negative sunspot N and the negative magnetic channel, displaying overall strong shearing motion over the core region of the flare.
These observations are suggestive that the flux cancellation ongoing along the PILs may be driven by the strong shearing motion and is tightly related to the
triggering of the flare and the subsequent formation of the filament.

\subsection{The Flare Associated with the Formation of a Sigmoid}
The light curves of AIA 131 \AA\ and \emph{GOES} 1-8 \AA\ SXR flux display multiple impulsive peaks in the course of the flare (Figure 1({\it h})).
This is similar to our previous observations \citep{yang19a} that each impulsive peak may correspond to a magnetic reconnection process.
Figures 2({\it a})-({\it d}) show the evolution of the flare associated with the formation of a sigmoid structure.
It is clear that a set of brighter and shorter hot loops, which apparently connect P1 to the negative magnetic channel,
firstly appeared near the negative sunspot N (panels ({\it a}) and ({\it b})). Afterwards, another set of brighter and shorter hot loops,
which connect the positive magnetic flux patches in between P and P1 to the negative magnetic channel, appeared close to the positive sunspot P (panel ({\it c})).
These hot loops appearing at different locations may be indicative of multiple magnetic reconnection processes occurring at different locations and different times.
As the flare progressed, longer hot loops apparently rooted in P and N were gradually formed and finally shown as a remarkable sigmoid structure (panels ({\it d}) and ({\it h})).
The sigmoid formed along the main PILs and had an inverse S-shape, which suggests negative, left-handed twist.

The evolution of the flare ribbons is shown on Figures 2({\it e})-({\it g}). The flare ribbons first appeared at the region where the negative magnetic channel
was in contact with P1 (panel ({\it e})). These ribbons were distributed on either side of the main PILs
and in line with the footpoints of the hot loops that appeared near N (panels ({\it a}) and ({\it e})).
Then, they developed parallel to the main PILs in opposite directions and formed a two ribbon flare (panel ({\it f})).
It is worth mentioning that a conjugate compact footpoint brightenings appeared at P and N (panel ({\it f})),
which were consistent with the footpoints of the formed sigmoid (panels ({\it d}) and ({\it h})).
Subsequently, more intense flare brightenings occurred at the footpoints of the hot loops that appeared close to P (panel ({\it g})).
At the same time, the area of the conjugate compact footpoint brightenings were further expanded.
It is evident from the inserted $H_{\alpha}$ images that the most pronounced flare ribbons correspond to the conjugate compact footpoint brightenings
and the footpoints of the hot loops. They showed up as four compact footpoint beightenings in the $H_{\alpha}$ observation at about 21:44:54 UT.
This flare was confined, the two main ribbons did not separate perpendicular to the PILs and there were no clear CME signatures.
These observations taken together with the photospheric observations introduced before are highly reminiscent of the tether-cutting reconnection scheme \citep{moore01,liu13,chen14,chen18}.
More precisely, our observations fit very well with the physical picture that a confined flare is triggered by the tether-cutting reconnection process.
Therefore, we argue that the confined flare associated with the formation of the sigmoid in the present study are the results of
a sequential tether-cutting reconnection processes occurred between the magnetic fields connecting P to the negative magnetic channel
and the magnetic fields connecting P1 and its nearby positive flux patches to N.
Moreover, these reconnection may be driven by the strong shearing motion during the flux emergence.

\subsection{Chromospheric Evaporation at the Conjugate Footpoint Brigthenings}
Chromospheric evaporation, first described by \citet{ne68}, refers to the process during a solar flare
that chromospheric materials are heated and then expand rapidly upward into the low-density corona,
thus fill up the coronal loops giving the hot and dense post-flare loops that can be visible at the EUV and SXR wavelengths.
This process occurs when the flare energy deposit in chromosphere by non-thermal electrons or thermal conduction exceeds what can be shed by radiative losses.
The blueshifts that correspond to plasma upflows in emission lines formed at flare temperatures provide strong evidence for chromospheric evaporation.
It is recognized that the evaporation is identified as explosive when emission lines formed at flare temperatures exhibit blueshifts
while emission lines formed at chromospheric and transition region temperatures exhibit redshifts; the evaporation is considered as gentle
when emission lines formed at all temperatures exhibit blueshifts \citep{fis85,bro09,mil09,chen10,li11,young13,tian15}.

Using spectroscopic observations from the \emph{Hinode}/EIS, chromospheric evaporation at the conjugate footpoint brightenings is well diagnosed.
Figures 3({\it a} \sbond {\it c}) show the spatial distribution of line intensity, Doppler velocity, and line width
for the \ion{Fe}{15} 284.16 \AA\ line in the impulsive phase of the flare. The conjugate footpoint brightenings are the sites of rapid chromospheric heating,
they are also shown as a pair of intense, compact brightenings in the \ion{Fe}{15} 284.16 \AA\ emission line (panel({\it a})).
It is evident from the Doppler velocity map (panel({\it b})) that the conjugate footpoint brightenings are dominated by blueshifts.
In particular, these blueshifts are accompanied by relatively large line widths (panel({\it c})).
We selected two points, which denoted by the crosses and marked ``1" and ``2",
to extract the \ion{He}{2} 256.32 \AA, \ion{Fe}{15} 284.16 \AA, and \ion{Fe}{16} 262.98 \AA\ line profiles.
Point 1 lies in the positive footpoint of the conjugate  footpoint brightenings while point 2 lies in the negative footpoint of the conjugate footpoint brightenings.
EIS scanned points 1 and 2 at 21:32:47 UT and 21:37:20 UT, respectively. The \ion{He}{2}, \ion{Fe}{15}, and \ion{Fe}{16} line profiles at point 1
are displayed on Figures 3({\it d} \sbond {\it f}) and at point 2 are displayed on Figures 3({\it g} \sbond {\it i}).
We fitted all the line profiles at points 1 and 2 using a single Gaussian function.
The fitting results are shown as red curves in Figures 3({\it d} \sbond {\it i}). It is seen that the chromospheric \ion{He}{2} line shows significant redshifts
while the coronal \ion{Fe}{15} and \ion{Fe}{16} lines show significant blueshifts at both locations.
The redshift velocities are measured to be 14.1 and 14.0 km s$^{-1}$ at points 1 and 2, respectively.
The blueshift velocities are measured to be -28.9 km s$^{-1}$ and -11.2 km s$^{-1}$  for the \ion{Fe}{15} line,
-30.3 km s$^{-1}$ and -17.6 km s$^{-1}$ for the  \ion{Fe}{16} line at points 1 and 2, respectively.
These results reveal that explosive evaporation occurred at the conjugate footpoint brightenings in the impulsive phase of the flare.

\subsection{Coronal Condensation Resulting in the Filament Formation}
The explosive evaporation occurring at the conjugate footpoint brightenings may provide heated material to make the formed sigmoid overdense,
which is conducive to trigger thermal instability or thermal non-equilibrium, causing catastrophic cooling and coronal condensation.
Figure 4 presents the formation and evolution of cool condensations in the sigmoid and the formation of a filament.
At 22:50:38 UT, about 34 minutes after the flare ended, a set of bright loops gradually appeared (panel ({\it a1})).
At the same time, the initial condensation (as indicated by the red arrow in panel ({\it b1})), which can be clearly identified in the AIA 335 \AA\ observation,
formed at about the middle portion of these bright loops. Thereafter, the condensations continued to grow and moved in opposite directions along these bright loops
( see panels ({\it a2} \sbond {\it a3}), ({\it b2} \sbond {\it b3}), and the associated animation), forming a long filamentary structure (panel ({\it c2} \sbond {\it c3})).
These condensations are short-lived. They finally moved toward and fallen rapidly to the footpoints of these bright loops.
As a result, the bright loops faded and the long filamentary structure disappeared.
Following the disappearance of the long filamentary structure, another set of bright loops, which has an overall inverse S-shape, gradually appeared (panels ({\it a3} \sbond {\it a4})).
Similarly, the cool condensations firstly formed at about the middle portion of these inverse S-shaped bright loops,
they then continued to grow and moved in opposite directions along these bright loops (panels ({\it a4}) and ({\it b4})),
resulting in the formation of two long, almost parallel filamentary structures (panel ({\it c4})). As the condensations persistently formed and moved in opposite directions along the two
filamentary structures, a filament with an inverse S-shape was formed (panels ({\it a5} \sbond {\it a6}), ({\it b5} \sbond {\it b6}), and ({\it c5} \sbond {\it c6})).
The formed filament was rooted in the locations corresponding to the conjugate footpoint brightenings (see Figure 1({\it c})),
with its positive ends anchored in the positive sunspot P and its negative ends anchored in the negative sunspot N (panel ({\it c6})).
The filament is sinistral, indicating negative, left-handed twist \citep{mar98}, which is in line with the sigmoid.

The gradual emergence of the bright loops should be the result of the gradual cooling of the formed sigmoid over time after the flare.
Apparently, the first appeared bright loops are low-lying relative to the inverse S-shaped bright loops.
The successive appearance of cool condensates in the low-lying bright loops and then inverse S-shaped bright loops
are consistent with the observation of \citet{liu12} and the simulation of \citet{kar08}. They implied that cool condensations firstly formed at
those low-lying loops are due to their shorter lengths and/or their greater densities from gravitational stratification.
To trace the dynamic evolution of the cool condensations in the low-lying and the inverse S-shaped bright loops, spacetime plots along slice ``AB" and ``CD" in Figures 4({\it a1}) and ({\it a5})
were constructed from AIA 335 \AA\ images, and the result was provided in Figures 5({\it c}) and ({\it d}).
The cool condensations moved in opposite directions along the low-lying and the inverse S-shaped bright loops are clearly shown on the spacetime plots.
we find that the cool condensations moved along the low-lying bright loops to their positive ends with a mean velocity of about 43.3 km s$^{-1}$
and to their negative ends with a mean velocity of about 39.4 km s$^{-1}$; the cool condensations moved along the inverse S-shaped bright loops to their positive ends
with a mean velocity of about 36.6 km s$^{-1}$ and to their negative ends with a mean velocity of about 37.1 km s$^{-1}$.

As the sigmoid gradually cooling over time after the flare, similar to the off-limb condensation events \citep{ber12,liu12,li18}
that the peak brightness of the appearing bright loops should progressively shift in time
from the AIA broadband channels with higher characteristic temperatures to the channels with lower characteristic temperatures.
However, this trend can not be definitively detected in this on-disk observation owing to intermittent magnetic activities still occurred after the flare.
It is interesting to note that a transient brightening (as denoted by the blue arrow in Figure 5({\it a}))
appeared in the inverse S-shaped loops during the formation of the filament.
It cooled down over time and finally formed cool condensations (as pointed out by the red arrow in Figure 5({\it b})) in the inverse S-shaped loops.
This transient feature is very similar to the long, transient brightening simulated by \citet {luna12}.
They suggested that this long, transient brightening is the consequence of the evaporation and condensation.
In the red dashed rectangle, enclosing the long, transient brightening in Figure 5({\it a}), the light curves of the AIA 211, 193, and 171 \AA\ channels were calculated
and displayed in Figure 5({\it e}) as blue, red, and green curves, respectively. It is obvious that the peak brightness of the long, transient brightening progressively shifts in time
from the 211 \AA\ channel, through the 193 \AA\ channel, to the 171 \AA\ channel. The long, transient brightening peaked first in the 211 \AA\ channel at about 23:21:36 UT.
With a delay of about 72 seconds, it peaked in the 193 \AA\ channel at about 23:22:48 UT. Then, it peaked in the 171 \AA\ channel at about 23:23:17 UT,
29 (101) seconds later than the peak of AIA 193 (211) \AA\ light curve. These results strongly suggest that hot plasma evaporated from the chromosphere cools down
and condenses in the sigmoid to form the cool dense filament, probably due to the onset of thermal instability or thermal non-equilibrium.

\section{Conclusion and Discussion}
In this study, we report a clear observation of a solar filament formed through chromospheric evaporation and subsequent coronal condensation after an M-class confined flare.
Our investigations show that this confined flare is triggered by a sequential tether-cutting reconnection processes as strong magnetic shear occurring during flux emergence in the NOAA AR 11967.
Owing to the tether-cutting reconnection, an elongated magnetic structure is newly created above the flaring PIL, which separately bridges its footpoints at two conjugate compact brightening regions.
The spectroscopic observations from the \emph{Hinode}/EIS reveal that explosive chromospheric evaporation takes place at the two compact brightening regions in the impulsive phase of the flare.
In this course, because the explosive injection of heated chromospheric plasma from its footpoints, the newborn elongated magnetic structure soon manifests as an X-ray sigmoid.
After the flare, cool material continuously condenses in the middle section of the sigmoid and then moves in opposite directions along the magnetic field lines of the sigmoid,
eventually resulting in the formation of a filament in the 304 \AA \ passband. This indicates that the evaporated hot plasma, which trapped in the sigmoid,
cools down to at least $\sim$ 0.05 MK within tens of minutes, possibly due to the onset of thermal instability or thermal non-equilibrium.
These observations demonstrate that magnetic reconnection not only directly create the desirable magnetic field structure for filament formation,
but also indirectly supply cold and dense material for the forming filament in tens of minutes through resultant chromospheric evaporation and subsequent coronal condensation.

Observational evidence of chromospheric evaporation has not been detected in filament formation events,
but has been documented in numerous flare events of different magnitude \citep{ding96,mil09,ning09,cheng15,li15,zhang16,huang20}.
In the framework of the standard flare model \citep{pri02}, electrons are accelerated at or near a magnetic reconnection site in the corona,
the accelerated electron beams propagate downward along newly reconnected magnetic field lines and then bombard the chromosphere to generate chromospheric evaporation.
Meanwhile, thermal conduction from the reconnection site may also contribute to the heating.
In our observation, in the course of the formation of the sigmoid by the sequential tether-cutting reconnection processes,
it is quite possible that reconnection-accelerated electron beams or thermal conduction from the reconnection sites
may travel or spread along the magnetic field lines of the sigmoid in opposite directions and toward their conjugate chromospheric footpoints.
As a result, the conjugate chromospheric footpoints of the sigmoid would be heated, manifesting as the conjugate compact footpoint brightenings.
Owing to the heating, as confirmed by the \emph{Hinode}/EIS spectroscopic observation that explosive evaporation occurred at the conjugate footpoint brightenings,
heated chromospheric plasma may thus be evaporated into the sigmoid and make it overdense.
Subsequently, thermal instability or thermal non-equilibrium might be triggered, causing catastrophic cooling and coronal condensation to form the cold, dense material of the filament.

The evaporation-condensation model is a promising candidate that accounts for the filament formation.
However, unambiguous observations directly validating this model have been missing up to now.
A handful of previous observations have shown that coronal condensation can lead to the formation of prominences \citep{ber12,liu12,viall20},
but the origin of the hot plasma involved in the coronal condensation has been questionable. In these observations and previous simulations \citep{ant91,kar08,luna12,xia16},
the formation of the filament magnetic structure and its cold dense material have been treated separately.
In actual fact, the formation of filaments are the results of both magnetic and thermal evolution.
\citet{kan17} proposed a reconnection-condensation model, which treated the formation of the filament magnetic structure and its cool dense material together,
to explain the formation of filaments. It is worth noting that any artificial localized heating in the chromosphere
is not considered in their model. Their model does not include the chromosphere and the resultant prominence mass only reaches the observed lower limit of typical prominence densities.
The magnetic reconnection in their model forms the filament magnetic structure and directly triggers thermal instability causing coronal condensation.
The reconnection induced topology changes triggering thermal instability and the formation of cool plasma condensations was subsequently confirmed in coronal rain observations \citep{li18,mas19,li21}.
In the present study, the formation of the filament magnetic structure and its cold dense material are treated together from the perspective of observation for the first time.
Our observations demonstrate that the formation of the filament magnetic structures by magnetic reconnection also has a profound influence on the subsequent thermal evolution occurring in it.
Different from \citet{kan17} and \citet{li18}, the magnetic reconnection in our observation might affect the subsequent thermal evolution in the filament magnetic structures
by heating their chromospheric footpoints during their formation.
This is consistent with the evaporation-condensation model \citep{ant91,ant00}.
This observation, for the first time, presents unambiguous observational evidence that both chromospheric evaporation
and subsequent coronal condensation take place in the formation of a filament, in support of the evaporative-condensation model.

The formed filament was short-lived. It only lived for about an hour and then gradually disappeared.
In the framework of the evaporation-condensation model, numerous simulations have shown that deep magnetic dips are necessary for supporting the cool condensations \citep{luna12,zhou20}.
In this scenario, a typical dynamic phenomenon that can usually be seen during the filament formation is the longitudinal oscillation of the filament \citep[see the review by][]{chen20a}.
In the absence of magnetic dips, the formation of long-lived filaments requires constant chromospheric footpoint heating to maintain a stable cycle,
in which plasma evaporates from the chromosphere, condenses into the filament in the corona, and drains back to the chromosphere along the magnetic field lines of the filament \citep{kar01}.
In the present study, any oscillations during the formation and disappearance of the filament are not observed.
Moreover, continual chromospheric footpoint heating and coronal condensation are not occurred after the filament formed.
It is likely that the filament might not contain deep magnetic dips, which are suitable for supporting the cool condensations.
The cool condensations could not be stably supported in the filament and thus drain in opposite directions along the filament, ultimately causing it to be short-lived.
The real and detailed magnetic field structure of the filament needs to be further investigated.

Generally, the formation of a quiescent prominence is days and even weeks long and the quiescent prominences are long-lived.
\citet{ber12} found that quiescent prominences can be formed via in situ condensation of hot plasma contained in the core of the
coronal cavity. However, the origin of the hot plasma is not yet identified. Although flux emergence and local reconnection have been suggested,
the possibility of chromospheric evaporation can not be ruled out. In the quiet Sun, the explosive events, such as reported in this paper, never happen.
But during the formation and evolution of the quiescent filaments, photospheric magnetic flux cancellation,
which might be closely related to the magnetic reconnection occurring in the lower atmosphere of the Sun, was found to be abundant near the PILs \citep{mar98,mac10}.
This process can also heat the footpoints of the reconnected coronal loops \citep{yang16,yang18}.
It is possible that gentle evaporation might be driven by the magnetic reconnection occurring in the lower atmosphere of the Sun during the formation of the magnetic structure of
the quiescent prominences. In addition, \citet{zhou20} recently proposed that turbulent heating on the solar surface can randomly evaporate materials from the solar surface to the corona,
leading to quiescent prominence formation. So far, observational evidence of chromospheric evaporation has not been observed during the formation and evolution of quiescent prominences.
The applicability of the evaporative-condensation model to quiescent prominences needs to be verfied by more high-resolution spectral and imaging observations in the future.
In the active region, the explosive events reported in this paper could frequently happen, and they will be more likely to be associated with active region filament formation.

\begin{acknowledgments}
We thank the referee for valuable suggestions and comments that greatly improved the paper.
The authors thank the scientific/engineering team of \emph{SDO}, \emph{Hinode}, \emph{GOES}, NVST, and GONG for providing the excellent data.
We would like to thank Dr. Hechao Chen for his fruitful comments.
This work is supported by the National Key R\&D Program of China (2019YFA0405000), the Natural Science Foundation of China, under grants 12073072,
11633008, 11933009, 12173084, 11873088, and 11703084; the CAS ``Light of West China" Program;
the Open Research Program of the Key Laboratory of Solar Activity of Chinese Academy of Sciences (KLSA202018);
the CAS grant ``QYZDJ-SSW-SLH012"; the Yunnan Natural Science Foundation, under grants 202001AU070052 and 202101AV070004;
and the Project Founded by China Postdoctoral Science Foundation 2020M671639.
\end{acknowledgments}

\newpage
\begin{figure}
\epsscale{1.}
\plotone{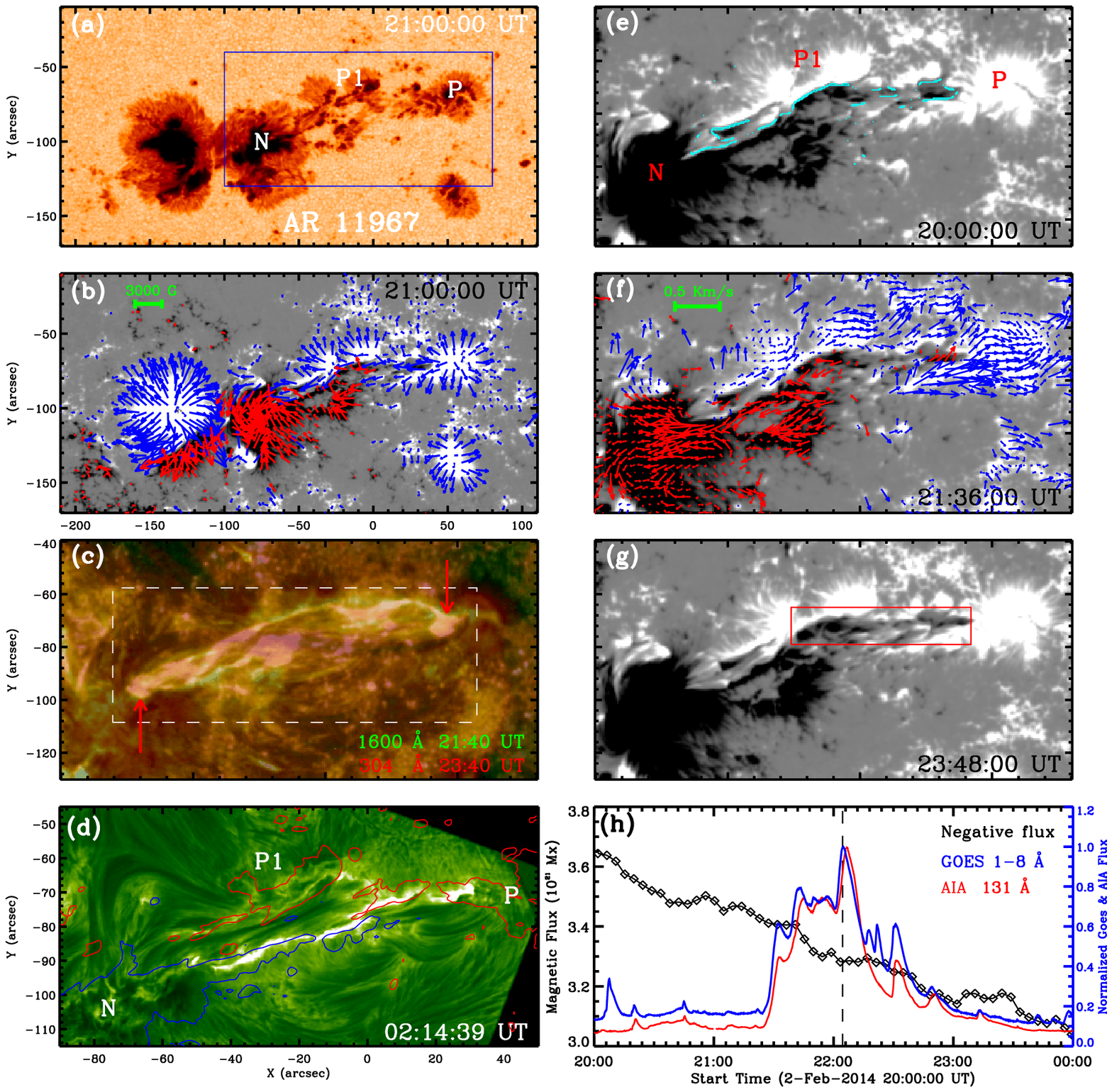}
\caption{({\it a}),({\it b}) A continuum intensity image and a vector magnetogram of AR 11967 taken by \emph{SDO}/HMI at 21:00 UT on 2014 February 2.
({\it c}) Composite image of the AIA 1600 \AA\ and 304 \AA\ passbands displaying the ribbons for the M1.3 flare and the formed filament.
The two red arrows point to the conjugate compact footpoint brightenings of the flare.
({\it d}) A NVST $H_{\alpha}$ image, along with HMI vertical magnetic field contours overplotted as red/blue at $\pm 800$  G levels.
({\it e})\sbond({\it g}) The vertical component of the HMI vector magnetic field showing the cancellation of opposite polarities in the core region of the flare.
The cyan curves outline the main PILs. The red (blue) arrows in panel ({\it b}) are the horizontal magnetic field vectors
and in  panel ({\it f}) are the tangential velocity vectors, which originate from negative (positive) longitudinal field.
The tangential velocity vectors are inferred from the DAVE4VM technique and averaged from 20:00 UT to 22:00 UT.
Labels ``P," and ``P1"  register the sunspots with positive polarity, while ``N" denotes the sunspot with negative polarity.
({\it h}) The normalized \emph{GOES} SXR flux at 1\sbond8 \AA\ , the normalized AIA 131 \AA\ light curves extracted from the dashed rectangle, and the unsigned negative magnetic
flux averaged over the red rectangle. The vertical dashed line marks the peak time of the flare.
The blue rectangle denotes the FOV of panels ({\it c}) and ({\it e}\sbond{\it g}).
}
\end{figure}

\begin{figure}
\epsscale{1.1}
\plotone{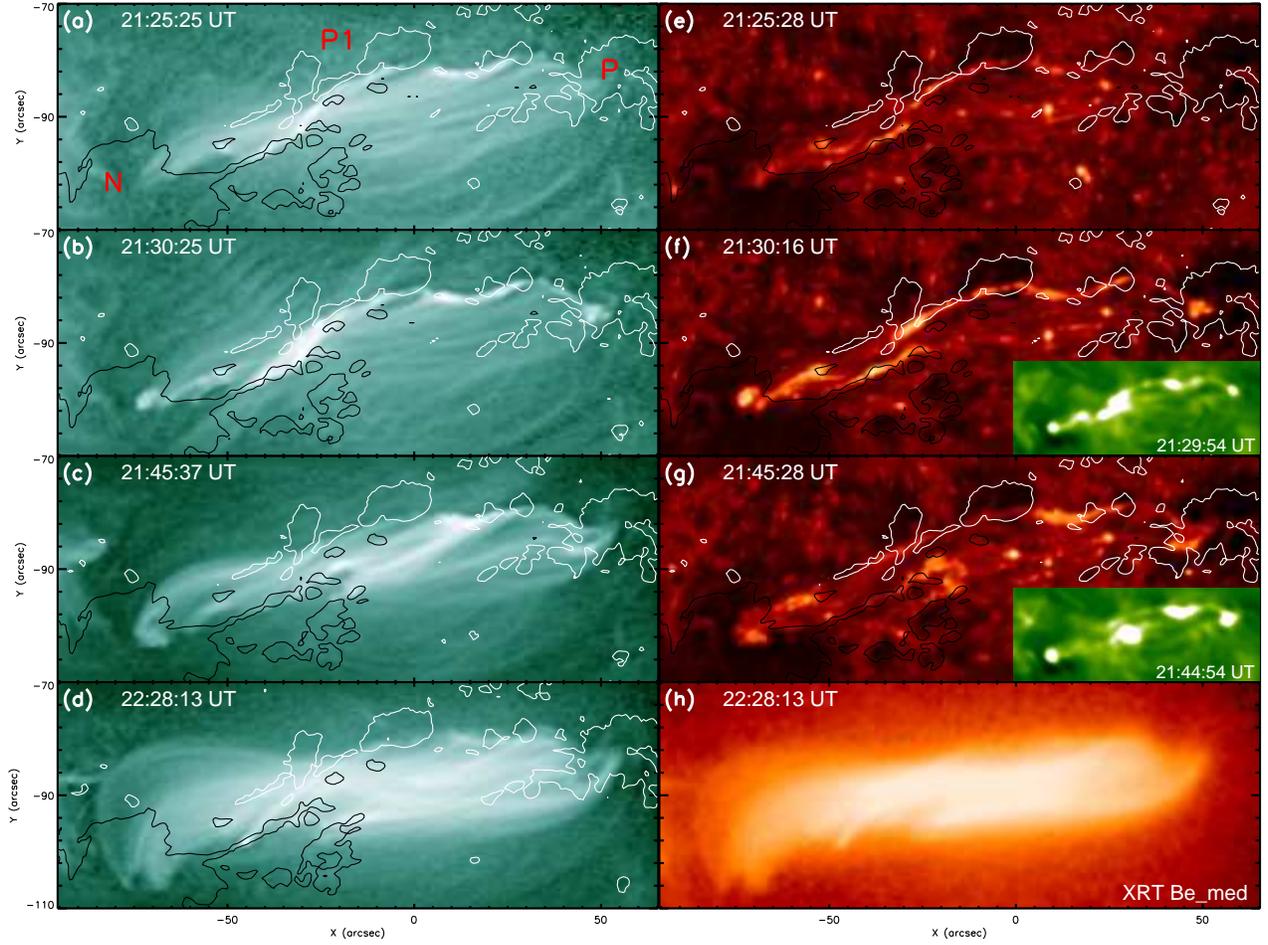}
\caption{\emph{SDO}/AIA 94 \AA\ (panels ({\it a}\sbond{\it d})) and \emph{Hinode}/XRT Be-med (panel({\it h})) images
showing the formation of a sigmoid structure. AIA 1600 \AA\ (panels ({\it e}\sbond{\it g})) and GONG $H_{\alpha}$ (the inserted images) images
displaying the evolution of the flare ribbons. Iso-Gauss contours of $\pm 800$ G are superposed by white and black curves on panels ({\it a}\sbond{\it g}).
An animation of panels ({\it a}\sbond{\it d}) and ({\it e}\sbond{\it g}) is available.
The animation spans from 21:00:37 UT to 22:39:52 UT on 2014 February 2, and the time cadence is 9 s.
(An animation of this figure is available.)}
\end{figure}

\begin{figure}
\epsscale{1.1}
\plotone{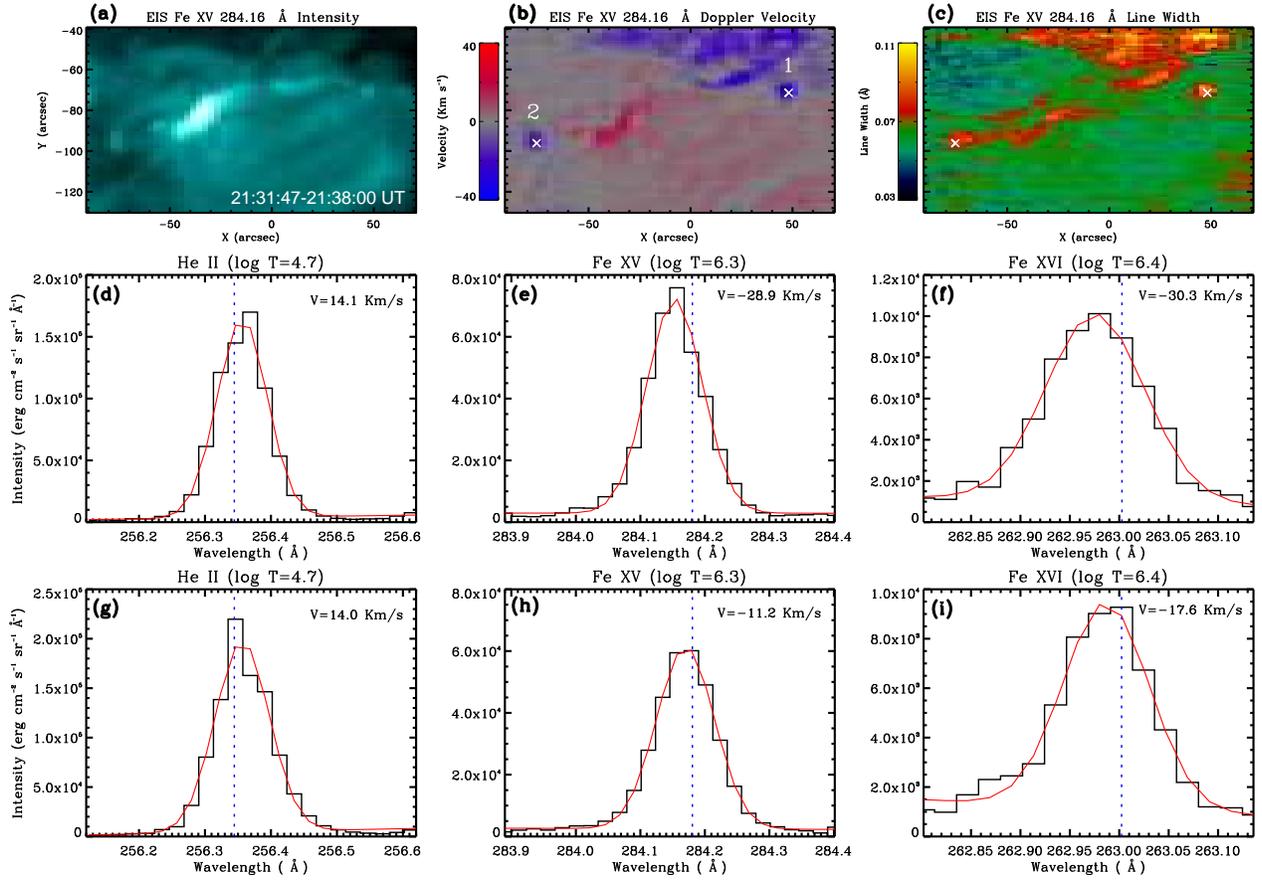}
\caption{({\it a} \sbond {\it c}) \emph{Hinode}/EIS \ion{Fe}{15} 284.16 \AA\ intensity, Doppler velocity, and line width maps.
The crosses marked ``1" and ``2"  are selected to perform a detailed spectral analysis.
Panels ({\it d} \sbond {\it f}) and ({\it g} \sbond {\it i}) are line profiles and fitting results for the \ion{He}{2}, \ion{Fe}{15}, and \ion{Fe}{16} lines
at point 1 and point 2, respectively. The histograms are observed profiles and the red curves are
fitting results. The dotted lines represent the rest wavelengths as measured from quiet-Sun regions.
}
\end{figure}

\begin{figure}
\epsscale{1.1}
\plotone{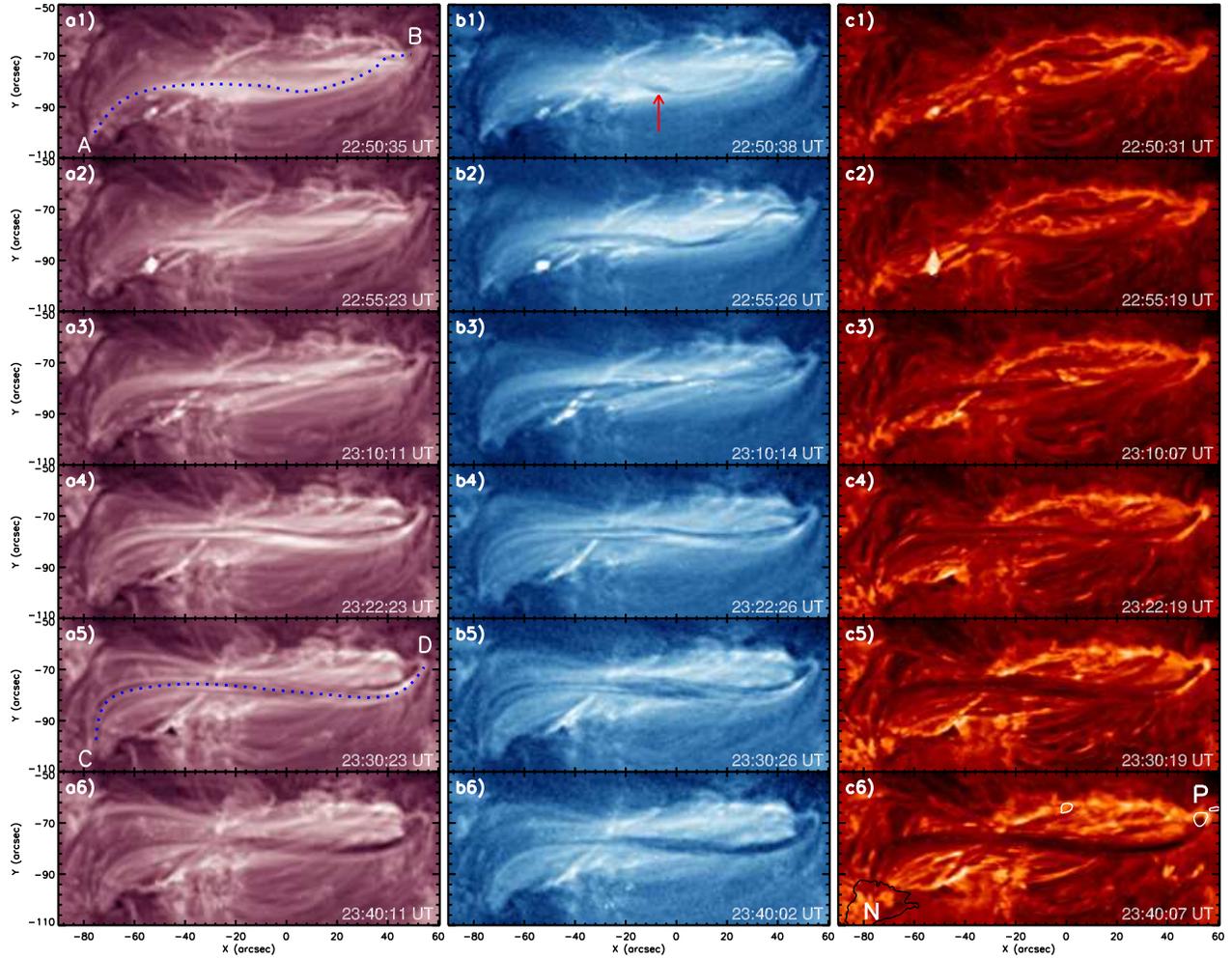}
\caption{Sequence of AIA 211 \AA\ (panels ({\it a1} \sbond {\it a6})), 335 \AA\ (panels ({\it b1} \sbond {\it b6})),
and 304 \AA\ (panels ({\it c1} \sbond {\it c6})) images present the formation of the filament by coronal condensation.
The red arrow points to the location where the initial condensation appeared.
The dashed lines ``AB" and ``CD"  mark the slit position of the time slices shown in Figures 5({\it c}) and ({\it d}), respectively.
Iso-Gauss contours of $\pm 2000$ G are superposed by white and black curves on panel ({\it c6}). An animation including the sequence of  AIA 304 \AA,
171 \AA, 193 \AA, 211 \AA, 335 \AA, 131 \AA, and 94 \AA\ images is available.
The animation spans from 22:40:01 UT to 23:59:59 UT on 2014 February 2, and the time cadence is 8 s.
(An animation of this figure is available.)}
\end{figure}

\begin{figure}
\epsscale{1.1}
\plotone{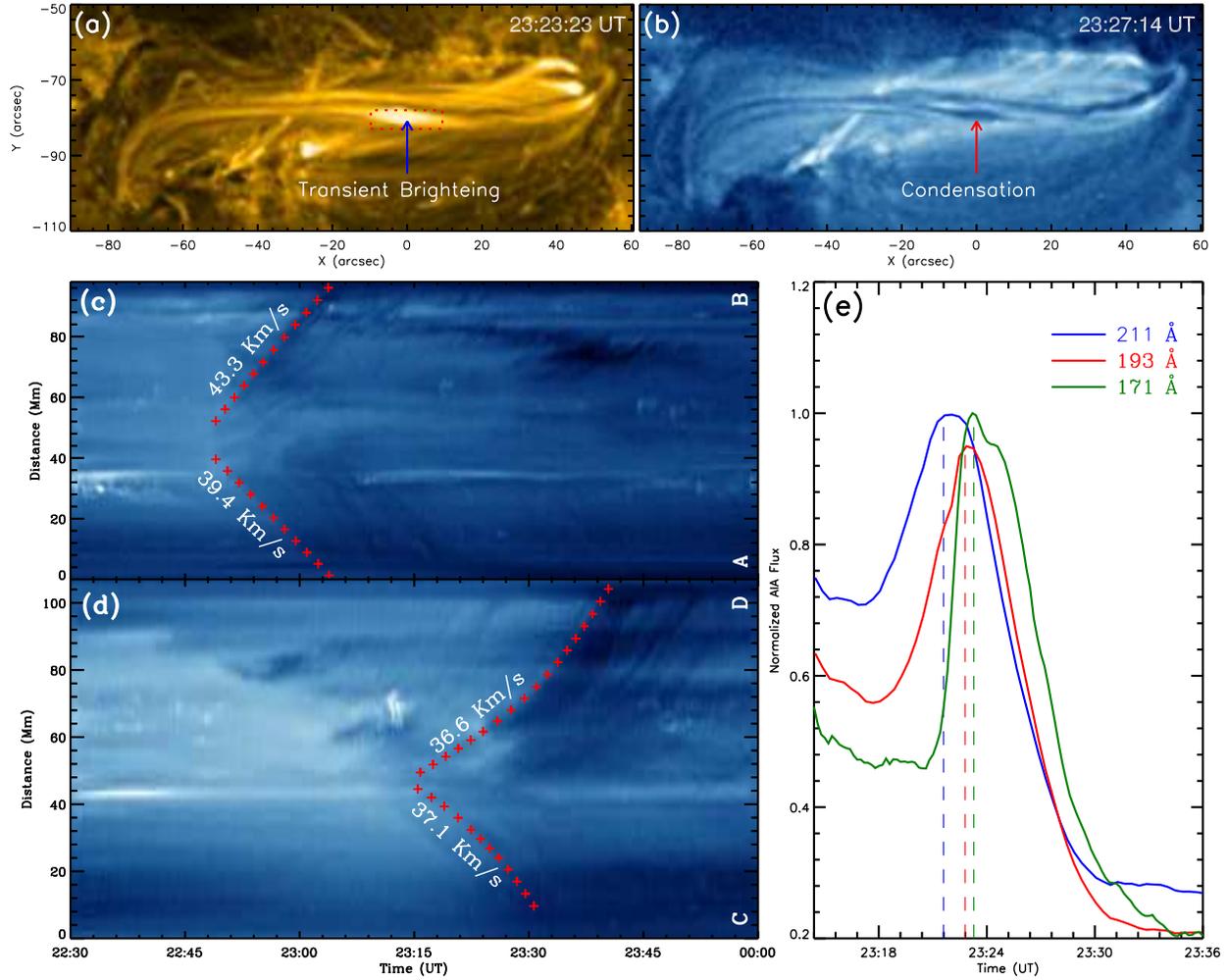}
\caption{AIA 171 \AA\ (panel ({\it a})) and  335 \AA\ (panel ({\it b})) images exhibit that a long, transient brightening cools to and condenses into the filament over time.
({\it c})\sbond({\it d}) Time-slice plots made from AIA 335 \AA\ images separately along the dashed lines AB and CD in Figure 4.
The red plus signs outline the motions of the cool condensations. ({\it e}) The normalized AIA light curves extracted from the red dashed rectangle.}
\end{figure}



\begin{thebibliography}{}
\bibitem[Antiochos \& Klimchuk(1991)]{ant91} Antiochos, S.~K. \& Klimchuk, J.~A.\ 1991, \apj, 378, 372. doi:10.1086/170437

\bibitem[Antiochos et al.(2000)]{ant00} Antiochos, S.~K., MacNeice, P.~J., \& Spicer, D.~S.\ 2000, \apj, 536, 494. doi:10.1086/308922

\bibitem[Anzer \& Heinzel(2005)]{anz05} Anzer, U. \& Heinzel, P.\ 2005, \apj, 622, 714. doi:10.1086/427817

\bibitem[Berger et al.(2012)]{ber12} Berger, T.~E., Liu, W., \& Low, B.~C.\ 2012, \apjl, 758, L37. doi:10.1088/2041-8205/758/2/L37

\bibitem[Berger et al.(2011)]{ber11} Berger, T., Testa, P., Hillier, A., et al.\ 2011, \nat, 472, 197. doi:10.1038/nature09925

\bibitem[Brosius(2009)]{bro09} Brosius, J.~W.\ 2009, \apj, 701, 1209. doi:10.1088/0004-637X/701/2/1209

\bibitem[Chae et al.(2001)]{chae01} Chae, J., Wang, H., Qiu, J., et al.\ 2001, \apj, 560, 476. doi:10.1086/322491

\bibitem[Chen \& Ding(2010)]{chen10} Chen, F. \& Ding, M.~D.\ 2010, \apj, 724, 640. doi:10.1088/0004-637X/724/1/640

\bibitem[Chen et al.(2018)]{chen18} Chen, H., Duan, Y., Yang, J., et al.\ 2018, \apj, 869, 78. doi:10.3847/1538-4357/aaead1

\bibitem[Chen et al.(2020b)]{chen20b} Chen, H., Hong, J., Yang, B., et al.\ 2020b, \apj, 902, 8. doi:10.3847/1538-4357/abb1c1

\bibitem[Chen et al.(2014)]{chen14} Chen, H., Zhang, J., Cheng, X., et al.\ 2014, \apjl, 797, L15. doi:10.1088/2041-8205/797/2/L15

\bibitem[Chen et al.(2020a)]{chen20a} Chen, P.-F., Xu, A.-A., \& Ding, M.-D.\ 2020a, Research in Astronomy and Astrophysics, 20, 166. doi:10.1088/1674-4527/20/10/166

\bibitem[Cheng et al.(2015)]{cheng15} Cheng, X., Hao, Q., Ding, M.~D., et al.\ 2015, \apj, 809, 46. doi:10.1088/0004-637X/809/1/46

\bibitem[Culhane et al.(2007)]{cul07} Culhane, J.~L., Harra, L.~K., James, A.~M., et al.\ 2007, \solphys, 243, 19. doi:10.1007/s01007-007-0293-1

\bibitem[DeVore \& Antiochos(2000)]{dev00} DeVore, C.~R. \& Antiochos, S.~K.\ 2000, \apj, 539, 954. doi:10.1086/309275

\bibitem[Ding et al.(1996)]{ding96} Ding, M.~D., Watanabe, T., Shibata, K., et al.\ 1996, \apj, 458, 391. doi:10.1086/176822

\bibitem[Engvold(1998)]{eng98} Engvold, O.\ 1998, IAU Colloq.~167: New Perspectives on Solar Prominences, 150, 23

\bibitem[Fan(2001)]{fan01} Fan, Y.\ 2001, \apjl, 554, L111. doi:10.1086/320935

\bibitem[Field(1965)]{fie65} Field, G.~B.\ 1965, \apj, 142, 531. doi:10.1086/148317

\bibitem[Fisher et al.(1985)]{fis85} Fisher, G.~H., Canfield, R.~C., \& McClymont, A.~N.\ 1985b, \apj, 289, 425. doi:10.1086/162902

\bibitem[Golub et al.(2007)]{gol07} Golub, L., Deluca, E., Austin, G., et al.\ 2007, \solphys, 243, 63. doi:10.1007/s11207-007-0182-1

\bibitem[Huang et al.(2021)]{huang21} Huang, C.~J., Guo, J.~H., Ni, Y.~W., et al.\ 2021, \apjl, 913, L8. doi:10.3847/2041-8213/abfbe0

\bibitem[Huang et al.(2020)]{huang20} Huang, Z., Zhang, Q., Xia, L., et al.\ 2020, \apj, 897, 113. doi:10.3847/1538-4357/ab96bd

\bibitem[Kamio et al.(2010)]{kam10} Kamio, S., Hara, H., Watanabe, T., et al.\ 2010, \solphys, 266, 209. doi:10.1007/s11207-010-9603-7

\bibitem[Kaneko \& Yokoyama(2017)]{kan17} Kaneko, T. \& Yokoyama, T.\ 2017, \apj, 845, 12. doi:10.3847/1538-4357/aa7d59

\bibitem[Karpen \& Antiochos(2008)]{kar08} Karpen, J.~T. \& Antiochos, S.~K.\ 2008, \apj, 676, 658. doi:10.1086/526335

\bibitem[Karpen et al.(2001)]{kar01} Karpen, J.~T., Antiochos, S.~K., Hohensee, M., et al.\ 2001, \apjl, 553, L85. doi:10.1086/320497

\bibitem[Keppens \& Xia(2014)]{kep14} Keppens, R. \& Xia, C.\ 2014, \apj, 789, 22. doi:10.1088/0004-637X/789/1/22

\bibitem[Kippenhahn \& Schl{\"u}ter(1957)]{ks57} Kippenhahn, R. \& Schl{\"u}ter, A.\ 1957, \zap, 43, 36

\bibitem[Kosugi et al.(2007)]{kos07} Kosugi, T., Matsuzaki, K., Sakao, T., et al.\ 2007, \solphys, 243, 3. doi:10.1007/s11207-007-9014-6

\bibitem[Kuperus \& Raadu(1974)]{kr74} Kuperus, M. \& Raadu, M.~A.\ 1974, \aap, 31, 189

\bibitem[Lemen et al.(2012)]{lem12} Lemen, J. R., Title, A. M., Akin, D. J., et al. 2012, \solphys, 275, 17

\bibitem[Leroy(1989)]{ler89} Leroy, J.~L.\ 1989, Dynamics and Structure of Quiescent Solar Prominences, 150, 77. doi:10.1007/978-94-009-3077-3\_13

\bibitem[Li et al.(2015)]{li15} Li, D., Ning, Z.~J., \& Zhang, Q.~M.\ 2015, \apj, 813, 59. doi:10.1088/0004-637X/813/1/59

\bibitem[Li et al.(2021)]{li21} Li, L., Peter, H., Chitta, L.~P., et al.\ 2021, \apj, 910, 82. doi:10.3847/1538-4357/abe537

\bibitem[Li et al.(2018)]{li18} Li, L., Zhang, J., Peter, H., et al.\ 2018, \apjl, 864, L4. doi:10.3847/2041-8213/aad90a

\bibitem[Li \& Ding(2011)]{li11} Li, Y. \& Ding, M.~D.\ 2011, \apj, 727, 98. doi:10.1088/0004-637X/727/2/98

\bibitem[Liu et al.(2013)]{liu13} Liu, C., Deng, N., Lee, J., et al.\ 2013, \apjl, 778, L36. doi:10.1088/2041-8205/778/2/L36

\bibitem[Liu et al.(2016)]{liu16} Liu, R., Chen, J., Wang, Y., et al.\ 2016, Scientific Reports, 6, 34021. doi:10.1038/srep34021

\bibitem[Liu et al.(2012)]{liu12} Liu, W., Berger, T.~E., \& Low, B.~C.\ 2012, \apjl, 745, L21. doi:10.1088/2041-8205/745/2/L21

\bibitem[Liu et al.(2005)]{liu05} Liu, Y., Kurokawa, H., \& Shibata, K.\ 2005, \apjl, 631, L93. doi:10.1086/496919

\bibitem[Liu et al.(2014)]{liu14} Liu, Z., Xu, J., Gu, B.-Z., et al.\ 2014, Research in Astronomy and Astrophysics, 14, 705-718. doi:10.1088/1674-4527/14/6/009

\bibitem[Lites et al.(2010)]{lit10} Lites, B.~W., Kubo, M., Berger, T., et al.\ 2010, \apj, 718, 474. doi:10.1088/0004-637X/718/1/474

\bibitem[Luna et al.(2012)]{luna12} Luna, M., Karpen, J.~T., \& DeVore, C.~R.\ 2012, \apj, 746, 30. doi:10.1088/0004-637X/746/1/30

\bibitem[Mackay et al.(2010)]{mac10} Mackay, D.~H., Karpen, J.~T., Ballester, J.~L., et al.\ 2010, \ssr, 151, 333. doi:10.1007/s11214-010-9628-0

\bibitem[Martens \& Zwaan(2001)]{mar01} Martens, P.~C. \& Zwaan, C.\ 2001, \apj, 558, 872. doi:10.1086/322279

\bibitem[Martin(1998)]{mar98} Martin, S.~F.\ 1998, \solphys, 182, 107. doi:10.1023/A:1005026814076

\bibitem[Mason et al.(2019)]{mas19} Mason, E.~I., Antiochos, S.~K., \& Viall, N.~M.\ 2019, \apjl, 874, L33. doi:10.3847/2041-8213/ab0c5d

\bibitem[Milligan \& Dennis(2009)]{mil09} Milligan, R.~O. \& Dennis, B.~R.\ 2009, \apj, 699, 968. doi:10.1088/0004-637X/699/2/968

\bibitem[Moore et al.(2001)]{moore01} Moore, R.~L., Sterling, A.~C., Hudson, H.~S., et al.\ 2001, \apj, 552, 833. doi:10.1086/320559

\bibitem[Neupert(1968)]{ne68} Neupert, W.~M.\ 1968, \apjl, 153, L59. doi:10.1086/180220

\bibitem[Ning et al.(2009)]{ning09} Ning, Z., Cao, W., Huang, J., et al.\ 2009, \apj, 699, 15. doi:10.1088/0004-637X/699/1/15

\bibitem[Okamoto et al.(2008)]{oka08} Okamoto, T.~J., Tsuneta, S., Lites, B.~W., et al.\ 2008, \apjl, 673, L215. doi:10.1086/528792

\bibitem[Orozco Su{\'a}rez et al.(2014)]{oro14} Orozco Su{\'a}rez, D., Asensio Ramos, A., \& Trujillo Bueno, J.\ 2014, \aap, 566, A46. doi:10.1051/0004-6361/201322903

\bibitem[Parker(1953)]{par53} Parker, E.~N.\ 1953, \apj, 117, 431. doi:10.1086/145707

\bibitem[Pesnell et al.(2012)]{pes12} Pesnell, W.~D., Thompson, B.~J., \& Chamberlin, P.~C.\ 2012, \solphys, 275, 3. doi:10.1007/s11207-011-9841-3

\bibitem[Priest \& Forbes(2002)]{pri02} Priest, E.~R. \& Forbes, T.~G.\ 2002, \aapr, 10, 313. doi:10.1007/s001590100013

\bibitem[Rust \& Kumar(1994)]{rus94} Rust, D.~M. \& Kumar, A.\ 1994, \solphys, 155, 69. doi:10.1007/BF00670732

\bibitem[Schmieder et al.(2014)]{schm14} Schmieder, B., Tian, H., Kucera, T., et al.\ 2014, \aap, 569, A85. doi:10.1051/0004-6361/201423922

\bibitem[Schou et al.(2012)]{sch12} Schou, J., Scherrer, P. H., Bush, R. I., et al. 2012, \solphys, 275, 229

\bibitem[Schuck(2008)]{sch08} Schuck, P.~W.\ 2008, \apj, 683, 1134. doi:10.1086/589434

\bibitem[Shen et al.(2019)]{shen19} Shen, Y., Qu, Z., Yuan, D., et al.\ 2019, \apj, 883, 104. doi:10.3847/1538-4357/ab3a4d

\bibitem[Song et al.(2017)]{song17} Song, H.~Q., Chen, Y., Li, B., et al.\ 2017, \apjl, 836, L11. doi:10.3847/2041-8213/aa5d54

\bibitem[Tian et al.(2015)]{tian15} Tian, H., Young, P.~R., Reeves, K.~K., et al.\ 2015, \apj, 811, 139. doi:10.1088/0004-637X/811/2/139

\bibitem[van Ballegooijen \& Martens(1989)]{van89} van Ballegooijen, A.~A. \& Martens, P.~C.~H.\ 1989, \apj, 343, 971. doi:10.1086/167766

\bibitem[Viall et al.(2020)]{viall20} Viall, N.~M., Kucera, T.~A., \& Karpen, J.~T.\ 2020, \apj, 905, 15. doi:10.3847/1538-4357/abc419

\bibitem[Wang et al.(2018)]{wang18} Wang, J., Yan, X., Qu, Z., et al.\ 2018, \apj, 863, 180. doi:10.3847/1538-4357/aad187

\bibitem[Wang et al.(2007)]{wangj07} Wang, J., Zhang, Y., Zhou, G., et al.\ 2007, \solphys, 244, 75. doi:10.1007/s11207-007-9038-y

\bibitem[Wang(1999)]{wang99} Wang, Y.-M.\ 1999, \apjl, 520, L71. doi:10.1086/312149

\bibitem[Wei et al.(2020)]{wei20} Wei, H., Huang, Z., Hou, Z., et al.\ 2020, \mnras, 498, L104. doi:10.1093/mnrasl/slaa134

\bibitem[Xia et al.(2011)]{xia11} Xia, C., Chen, P.~F., Keppens, R., et al.\ 2011, \apj, 737, 27. doi:10.1088/0004-637X/737/1/27

\bibitem[Xia \& Keppens(2016)]{xia16} Xia, C. \& Keppens, R.\ 2016, \apj, 823, 22. doi:10.3847/0004-637X/823/1/22

\bibitem[Xu et al.(2012)]{xu12} Xu, Z., Lagg, A., Solanki, S., et al.\ 2012, \apj, 749, 138. doi:10.1088/0004-637X/749/2/138

\bibitem[Yan et al.(2017)]{yan17} Yan, X.~L., Jiang, C.~W., Xue, Z.~K., et al.\ 2017, \apj, 845, 18. doi:10.3847/1538-4357/aa7e29

\bibitem[Yan et al.(2016)]{yan16} Yan, X.~L., Priest, E.~R., Guo, Q.~L., et al.\ 2016, \apj, 832, 23. doi:10.3847/0004-637X/832/1/23

\bibitem[Yang \& Chen(2019)]{yang19a} Yang, B. \& Chen, H.\ 2019, \apj, 874, 96. doi:10.3847/1538-4357/ab0c9e

\bibitem[Yang et al.(2018)]{yang18} Yang, B., Yang, J., Bi, Y., et al.\ 2018, \apj, 861, 135. doi:10.3847/1538-4357/aac37f

\bibitem[Yang et al.(2019)]{yang19b} Yang, B., Yang, J., Bi, Y., et al.\ 2019, \apj, 887, 220. doi:10.3847/1538-4357/ab557e

\bibitem[Yang et al.(2015)]{yang15} Yang, B., Jiang, Y., Yang, J., et al.\ 2015, \apj, 803, 86. doi:10.1088/0004-637X/803/2/86

\bibitem[Yang et al.(2016)]{yang16} Yang, B., Jiang, Y., Yang, J., et al.\ 2016, \apj, 816, 41. doi:10.3847/0004-637X/816/1/41

\bibitem[Young et al.(2013)]{young13} Young, P.~R., Doschek, G.~A., Warren, H.~P., et al.\ 2013, \apj, 766, 127. doi:10.1088/0004-637X/766/2/127

\bibitem[Zhang et al.(2016)]{zhang16} Zhang, Q.~M., Li, D., Ning, Z.~J., et al.\ 2016, \apj, 827, 27. doi:10.3847/0004-637X/827/1/27

\bibitem[Zhao et al.(2017)]{zhao17} Zhao, X., Xia, C., Keppens, R., et al.\ 2017, \apj, 841, 106. doi:10.3847/1538-4357/aa7142

\bibitem[Zhou et al.(2020)]{zhou20} Zhou, Y.~H., Chen, P.~F., Hong, J., et al.\ 2020, Nature Astronomy, 4, 994. doi:10.1038/s41550-020-1094-3

\end{thebibliography}
\end{document}